\documentclass{llncs} %

\usepackage{float}
\usepackage{graphicx}
\usepackage{wrapfig}

\usepackage[labelfont=bf]{caption}
\usepackage{color,soul}
\usepackage{amsmath} 
\usepackage{mathtools} 
\usepackage{amsfonts} 
\usepackage{amssymb}

\usepackage{array}
\usepackage{booktabs}
\usepackage{rotating}
\usepackage{multirow}
\usepackage{multicol}
\usepackage{lscape}
\usepackage{subfig}

\usepackage{wrapfig}
\usepackage{rotating}
\usepackage{epstopdf}

\usepackage[export]{adjustbox}

\usepackage{algorithm,algorithmicx,algpseudocode}

\usepackage{lineno}
\usepackage{soul}
\usepackage{color}
\usepackage{xcolor}
\usepackage{xargs}

\definecolor{brilliantrose}{rgb}{1.0, 0.33, 0.64}
\definecolor{capri}{rgb}{0.0, 0.75, 1.0}
	
\usepackage[colorlinks=true,linkcolor=capri,citecolor=capri]{hyperref}

\usepackage[colorinlistoftodos,prependcaption,textsize=tiny]{todonotes}
\newcommandx{\ISLEM}[2][1=]{\todo[linecolor=brilliantrose,backgroundcolor=brilliantrose!25,bordercolor=brilliantrose,#1]{#2}}

\usepackage{tikz,xcolor,hyperref}

\definecolor{lime}{HTML}{A6CE39}
\DeclareRobustCommand{\orcidicon}{
	\begin{tikzpicture}
	\draw[lime, fill=lime] (0,0) 
	circle [radius=0.16] 
	node[white] {{\fontfamily{qag}\selectfont \tiny ID}};
	\draw[white, fill=white] (-0.0625,0.095) 
	circle [radius=0.007];
	\end{tikzpicture}
	\hspace{-2mm}
}

\foreach \x in {A, ..., Z}{\expandafter\xdef\csname orcid\x\endcsname{\noexpand\href{https://orcid.org/\csname orcidauthor\x\endcsname}
			{\noexpand\orcidicon}}
}
			
\definecolor{darkgreen}{rgb}{0.53, 0.66, 0.42}

\begin{document}

\title{Adversarial Brain Multiplex Prediction From a Single Network for High-Order Connectional Gender-Specific Brain Mapping}

\titlerunning{Short Title}  

\author{Ahmed Nebli\orcidB{} \index{Nebli, Ahmed}\inst{1,2}  \and Islem Rekik\orcidA{} \index{Rekik, Islem}\inst{1}\thanks{ {corresponding author: irekik@itu.edu.tr, \url{http://basira-lab.com}, GitHub: \url{http://github.com/basiralab}. } This work is accepted for publication in the PRedictive Intelligence in MEdicine (PRIME) workshop Springer proceedings in conjunction with MICCAI 2020.}}

\institute{$^{1}$ BASIRA Lab, Faculty of Computer and Informatics, Istanbul Technical University, Istanbul, Turkey \\ $^{2}$ National School for Computer Science (ENSI), Mannouba, Tunisia }

\authorrunning{A Nebli et al.}

\maketitle              

\begin{abstract} 
Brain connectivity networks, derived from magnetic resonance imaging  (MRI), non-invasively quantify the relationship in function, structure, and morphology between two brain regions of interest (ROIs) and give insights into gender-related connectional differences. However, to the best of our knowledge, studies on gender differences in brain connectivity were limited to investigating \emph{pairwise} (i.e., \emph{low-order}) relationship ROIs, overlooking the complex \emph{high-order} interconnectedness of the brain as a network. A few recent works on neurological disorder diagnosis addressed this limitation by introducing \emph{the brain multiplex}, which in its shallow form, is composed of a source network intra-layer, a target intra-layer, and a convolutional inter-layer capturing the \emph{high-level} relationship between both intra-layers. However, brain multiplexes are built from at least two different brain networks, inhibiting its application to connectomic datasets with single brain networks such as functional networks. To fill this gap, we propose the first work on \emph{predicting brain multiplexes from a source network} to investigate gender differences. Recently, generative adversarial networks (GANs) submerged the field of medical data synthesis. However, although conventional GANs work well on \emph{images}, they cannot handle brain networks due to their non-Euclidean topological structure.  Differently, in this paper, we tap into the nascent field of \emph{geometric-GANs} (G-GAN) to design \emph{a deep multiplex prediction architecture} comprising (i) a geometric source to target network translator mimicking a U-Net architecture with skip connections and (ii) a conditional discriminator which classifies predicted target intra-layers by conditioning on the multiplex source intra-layers. Such architecture simultaneously learns the latent source network representation and the deep non-linear mapping from the source to target multiplex intra-layers. Our experiments on a large dataset demonstrated that predicted multiplexes significantly boost gender classification accuracy compared with source networks and identifies both low and high-order gender-specific multiplex connections.

\end{abstract}

\keywords{Geometric-generative adversarial networks $\cdot$ Brain multiplex prediction $\cdot$ Graph convolutional neural network  $\cdot$ Graph translation $\cdot$  Gender differences}

\section{Introduction}

The brain is a complex interconnected network encoding the connectional fingerprint of each individual and representing a notable biomarker of its gender \cite{Gong:2011}. In fact, several studies suggest that gender is a result of a natural distinction caused by human genetics and translated into dimorphic cortical connectivities. For instance, \cite{shirao2005} showed that males excel in memory and motor tasks while females have better linguistic and emotional processing yet more prone to anxiety and depression \cite{saunders1993}. Therefore, an accurate gender classification of brain networks might help better spot gender-biased brain disorders  and contribute to more reliable and personalized treatments. Despite the breadth of research on gender differences in brain functional and structural connectivity \cite{Gong:2011} as well as morphological connectivity \cite{Nebli:2019}, existing works are limited to investigating \emph{pairwise} (i.e., \emph{low-order}) relationship between ROIs, overlooking not only the complex \emph{high-order} interconnectedness of the brain as a network but also the topological configuration of brain ROIs connectivities.

To address the limitation of conventional low-order brain network representation, recent pioneering works \cite{raeper2018a,mahjoub2018} introduced the concept of a \emph{brain multiplex}, which in its shallow form, is composed of a source network intra-layer, a target intra-layer, and a convolutional inter-layer capturing the \emph{high-order} relationship between both intra-layers. Basically, a brain multiplex can be viewed as a tensor stacking two brain networks (also called intra-layers) and one inter-layer that encodes the similarity between these two intra-layers. While the intra-layers capture the \emph{low-order} pairwise relationship between ROIs, the inter-layer models the \emph{high-order} relationship between both intra-layer networks. In particular, the inter-layer is estimated by convolving one intra-layer with the other. The multiplex representation of brain connectivities boosted the classification accuracy in patients with early mild cognitive impairment \cite{raeper2018a} and patients with late mild cognitive impairment compared to the low-order brain network as well as their simple concatenation \cite{mahjoub2018}. Although compelling, building a brain multiplex depends on the availability of different intra-layer brain networks \cite{raeper2018a,mahjoub2018}, limiting its applicability to only multi-view (or multi-modal) brain connectomic datasets. One way to circumvent the issue of connectomic data with missing observations is to discard those samples, however this might not be convenient for devising learning-based classification frameworks. Ideally, one would learn how to predict a multiplex from a single intra-layer network (i.e., a source network) to (i) boost classification results, and (ii) discover low-order and high-order connectional biomarkers for the target classification task.

In recent years, the in-vogue generative adversarial networks (GANs) submerged the field of medical data synthesis \cite{yi2018} with unprecedented results in learning how to generate a target brain imaging modality from a different modality (e.g., computed tomography (CT) from MRI or T1-w from T2-w imaging) \cite{yi2018}. Despite the high efficiency of GANs in solving generative-related problems, all these methods focused on generating  \emph{only images}. However, brain connectivities as well as manifolds and graphs are essentially non-Euclidean spaces. To fill this gap in graph deep learning, \emph{geometric} deep leaning has been recently proposed to handle such data \cite{bronstein2017,guo2018} and propose new graph convolution and deconvolution operations. Given the absence of works on predicting brain multiplex as well as investigating the discriminative potential of these high-order brain connectional representations for gender classification, we design in this paper a deep brain multiplex prediction (DBMP) architecture comprising (i) a geometric source to target network translator mimicking a U-Net architecture with skip connections and (ii) a conditional discriminator which classifies predicted target intra-layers by conditioning on the multiplex source intra-layers. Such architecture \emph{simultaneously} learns the latent source network representation and the deep non-linear mapping from the source to target multiplex intra-layers. Taking into account its insensitivity to overtraining and its notable performance in paired classification, we train a support vector machine  (SVM) classifier in combination with a feature selection step on the predicted brain multiplexes for gender classification and connectional gender marker discovery.  The main innovative contributions of our work on both methodological and clinical levels are listed below.

\textbf{Methodological advance.} We propose the first geometric deep learning on \emph{brain multiplex prediction} based on G-GAN. This allows to \emph{learn} the multiplex instead of predefining it for each individual in the population and circumvents the scarcity of multi-view connectomic brain data compared to the breadth of computer vision datasets. Furthermore, to the best of our knowledge, this is the first work adapting geometric deep learning for brain connectome synthesis and classification in general.

\textbf{Clinical/scientific advances.} This is the first attempt to explore gender differences using brain multiplexes and reveal new connectional gender markers on both low and high-order levels. This contribution will enable early and personalized treatments for gender-biased brain disorders.

\begin{sidewaysfigure}
\centering
\includegraphics[width=19.5cm]{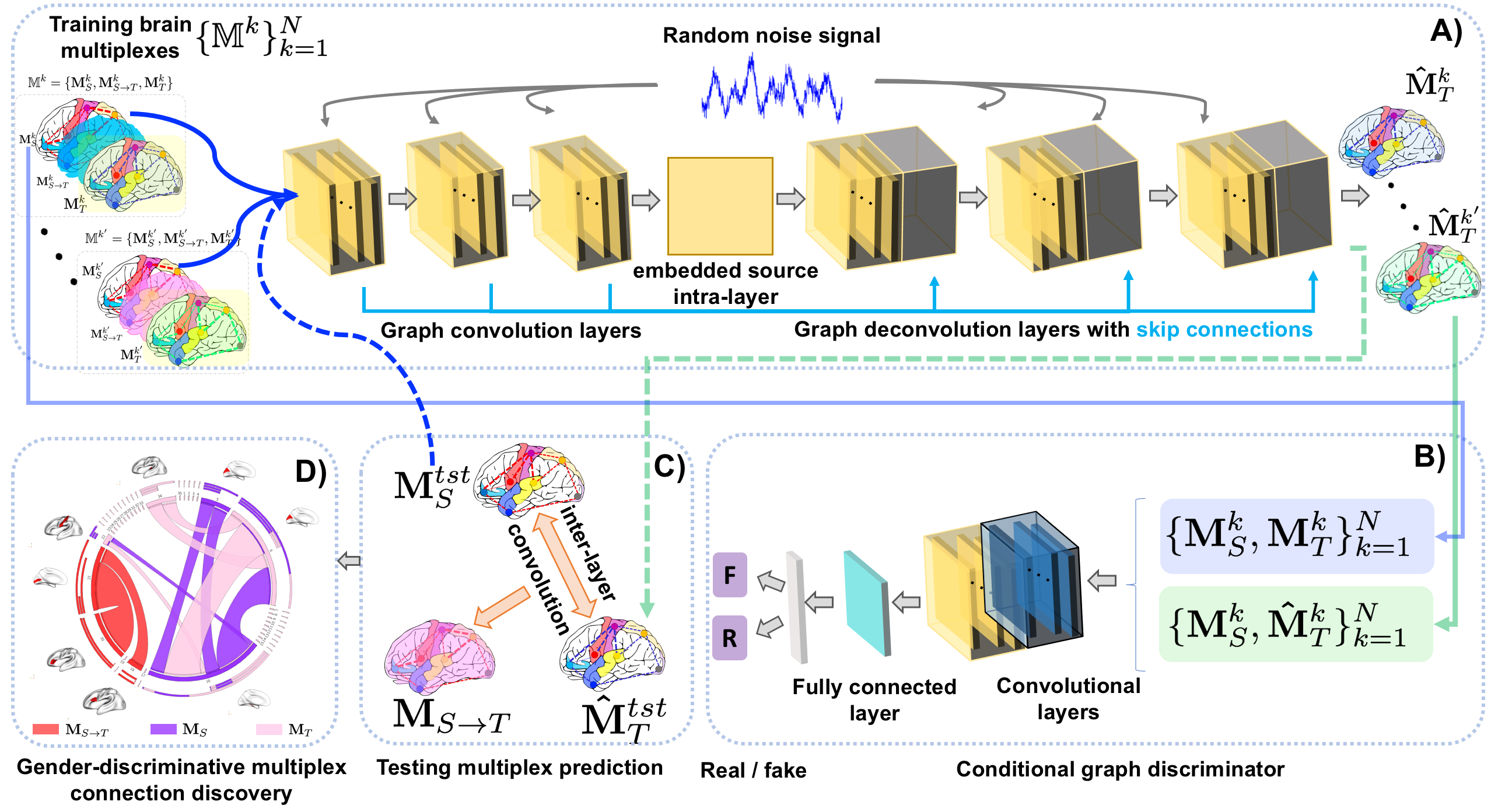}
\caption{\emph{Proposed deep brain multiplex prediction (DBMP) framework from a source network for gender classification and fingerprinting.} \textbf{(A)} Given a set of morphological brain multiplexes, we train a brain network translator to learn how to translate a source multiplex intra-layer $\mathbf{M}_S$ to the target intra-layer $\mathbf{M}_T$ . The translator architecture mimics that of U-Net comprising encoding and decoding layers with additional skip connections to pass over sample-specific representations to the encoding layers. \textbf{(B)} The conditional network discriminator aims to provide the translator with useful information about the multiplex generation quality. To do so,  for each of the $k$ training subjects, the conditional network discriminator inputs two sets of pairs of multiplex layers: $ \{ \mathbf{M}_S^k, \mathbf{M}_T^k \}_{k=1}^N$ and $ \{ \mathbf{M}_S^k, \mathbf{\hat{M}}_T^k \}_{k=1}^N$ and outputs  one value between $0$ and $1$ representing the \emph{realness} of the translator's predicted multiplex $\mathbf{\hat{M}}_T^k$.  \textbf{(C)} We produce the final multiplex $\mathbb{M}^k = \{\mathbf{M}_S^k, \mathbf{\hat{M}}_{S \rightarrow T}^k , \mathbf{\hat{M}}_T^k \}$ by stacking the source intra-layer, the predicted intra-layer, and inter-layer which is the result of a convolving both intra-layers. \textbf{(D)} We use an infinite feature selection strategy that selects the most gender discriminative brain multiplex connections to train a support vector machine classifier to assign each subject's brain multiplex gender. We display the most discriminative brain multiplex connections which belong to both intra-layers (source derived from mean average curvature and target from mean cortical thickness) as well as the high-order inter-layer. }
\label{fig:1}
\end{sidewaysfigure}

\section{Proposed Deep Brain Multiplex Prediction using G-GAN for Gender Fingerprinting}

In this section, we introduce the different steps of deep multiplex prediction framework from a source network for connectional gender fingerprinting. The proposed DBMP framework draws inspiration from \cite{guo2018}, which has pioneered the concept of graph or network translator. In typical GAN models, network generation is designed only for learning the distribution of network representations whereas network translation learns not only the latent network representation but also the generic translation mapping from source network to the target network simultaneously \cite{guo2018}. \textbf{Fig.}~\ref{fig:1} illustrates the proposed pipeline for DBMP with application to gender fingerprinting comprising the following steps: 1) network translator training, 2) conditional network discriminator learning, 3) linear SVM classifier training with feature selection, and 4) connectional gender discriminative brain multiplex connections discovery.

\textbf{Conventional shallow brain multiplex construction.} Given $N$ training subjects, let {$ \{ \mathbf{M}_S^k \}_{k=1}^N $} denote the set of $N$ source intra-layers and {$ \{ \mathbf{M}_T^k \}_{k=1}^N $} the set of $N$ target intra-layers. A conventional shallow brain multiplex  {$  \mathbb{M}^k $} for subject $k$  is defined as a tensor stacking three layers: a source intra-layer  {$  \mathbf{M}^k_{S} $}, a target intra-layer {$ \mathbf{M}^k_{T}$}, and a convolutional inter-layer {$ \mathbf{M}_{S \rightarrow T}^{k}$} encoding the similarity between the source and the target intra-layers \cite{raeper2018a}. The inter-layer is typically generated by convolving both multiplex intra-layers as follows \cite{mahjoub2018,raeper2018a}:

\begin{gather}
\mathbf{M}_{S \rightarrow T}^{k} (a,b) = \sum_{p} \sum_{q} \mathbf{M}^k_{S}(p,q)  \mathbf{M}^k_{T}(a - p +1, b-q+1) 
\end{gather}

Where $a$ and $b$ respectively denote the row and the column of a specific element in the inter-layer while $p$ and $q$ respectively denote the row and column of a specific element in the intra-layers.

\textbf{Deep brain multiplex prediction using G-GAN.} For the target prediction task, we assume that each sample is represented by a ground-truth source network intra-layer. Our goal is to learn the target intra-layer from the source, then convolve both to ultimately synthesize a shallow brain multiplex. To do so, inspired by the graph translator model introduced in \cite{guo2018} proposing new graph convolution and deconvolution operations on graphs, we formulate our multiplex prediction task in the spirit of generative adversarial learning. Specifically, given an input ground-truth source intra-layer {$  \mathbf{M}^k_{S} $} and a random noise, we predict the target intra-layer {$\hat{ \mathbf{M}}^k_{T}$}  by learning a local and global source to target {$ \mathbf{M}^k_{T}$} mapping where ground-truth and predicted target intra-layers are enforced to share sparse patterns. This is achieved via an adversarial training of a \emph{network translator} learning to generate fake target data from input source data by which it tries to \textit{fool} the conditional discriminator aiming to learn to tell real and fake samples apart. The main advantage of such architecture is its adversarial training resulting in one network training the other in a bi-directional way. 

Given a brain multiplex $\mathbb{M}^k$ of subject $k$, let ${  \mathbf{M}^k_{S} = (V, E, \mathbf{W})}$ denote a source intra-layer, represented as a fully-connected (directed or undirected) graph. $V$ is a set of $n$ nodes, ${E }{\subseteq}{ V \times V}$ is the set of edges and the set of weights for their corresponding edges are encoded in matrix $\mathbf{W} \in R^{ n \times n}$, namely a weighted adjacency matrix. Let ${e(i,j) \in E}$ denote the edge from the node ${v_i \in V}$ to the node ${v_j \in V}$ and  $\mathbf{W}_{i,j} \in \mathbf{W}$ denotes the corresponding weight of the edge $e(i,j)$.  Analogously, ${ \mathbf{M}^k_{T} = (V', E', \mathbf{W}')}$ denote the target intra-layer in $\mathbb{M}^k$. The translation $T_r$  from source to target intra-layer mapping is defined as ${T_r: U, S \rightarrow T}$, where $U$ refers to the random noise and $S$ and $T$ respectively denote the domains of both source and the target intra-layers. The proposed G-GAN based DBMP framework aims to synthesize a fake target intra-layer $\mathbf{\hat{M}}^k_{T} = T(  \mathbf{M}^k_{S}, U)$ that mimics the real target intra-layer $ \mathbf{M}^k_{T}$ in its topological and structural properties by minimizing the following adversarial loss function over all $k$ samples:

\begin{gather}
\mathcal{L}(T_r,D) = \mathbb{E}_{  \mathbf{M}^k_{S},   \mathbf{M}^k_{T}} [log {D} (  \mathbf{M}^k_{T} |   \mathbf{M}^k_{S})] + \mathbb{E}_{  \mathbf{M}^k_{S}, U} [log( 1- D(T_r(  \mathbf{M}^k_{S}, U) |   \mathbf{M}^k_{S}))],
\end{gather}

where both $T_r$ and $D$ are trying to minimize the output of this loss function in an adversarial way. Since we translate the source into the target intra-layer which might lie on different manifolds, we enforce sparsely shared patterns across translated source intra-layer and ground-truth target intra-layer via $L_1$ regularization which can boost the optimization process. The updated loss function is defined as follows to estimate the mapping translator $T_r$:  

\begin{gather}
T_r^{*}=  arg min_{T_r}max_{D} \mathcal{L}(T_r,D)+ \mathcal{L}_{L_{1}}(T_r),
\end{gather}

Where $ \mathcal{L}_{L_{1}}(T_r) = \mathbb{E}_{  \mathbf{M}^k_{S},   \mathbf{M}^k_{T},U}[||   \mathbf{M}^k_{T} - T_r(  \mathbf{M}^k_{S},U)||_{1} $. This is solved using ADAM optimizer for learning the translator and discriminator mappings alternatingly.

\textbf{Source to target intra-layer translator.} As displayed in \textbf{Fig}~\ref{fig:1}--A, the geometric translator acts as a U-Net with skip connections based on graph convolution in the three encoding layers and deconvolution in the three decoding layers. Specifically, the encoder network comprises two edge-to-edge convolutional layers and one edge-to-node convolutional layer, while the decoder network is composed of a node-to-edge deconvolutional layer and two edge-to-edge deconvolutional layers.

\textbf{Conditional network discriminator (\textbf{Fig.}~\ref{fig:1}--C).} The conditional discriminator is trained by minimizing the sum of two losses: the loss between the target intra-layers {$\{  \mathbf{M}^k_{T} \}_{k=1}^N$}  and the source {$\{  \mathbf{M}^k_{S} \}_{k=1}^N$} and the loss between the predicted target multiplex intra-layers  $\{  \mathbf{\hat{M}}^k_{T} \}_{k=1}^N$ and ground-truth target intra-layers {$\{  \mathbf{M}^k_{T} \}_{k=1}^N$}.  In order to compute these losses separately, the conditional network discriminator inputs for training sets of pairs of multiplex layers: $ \{ \mathbf{M}_S^k, \mathbf{M}_T^k \}_{k=1}^N$ and $ \{ \mathbf{M}_T^k, \mathbf{\hat{M}}_T^k \}_{k=1}^N$ to distinguish the pair including the ground truth $\mathbf{M}_T^k$ from the predicted one $\mathbf{\hat{M}}_T^k$ by the translator for each training subject $k$. The discriminator architecture comprises four stacked layers organized in the following order: two edge-to-edge convolutional layers, one edge-to-node convolutional layer, and fully connected layer followed by a softmax layer to output real or fake label for each input target intra-layer.

\textbf{Brain multiplex prediction from source network and classification.} Given the predicted target intra-layer for a testing subject $tst$, we synthesize its brain multiplex $\mathbb{\hat{M}}^{tst} = \{ \mathbf{M}_S^{tst}, \mathbf{M}_{S \rightarrow \hat{T}}^{tst},  \mathbf{\hat{M}}_T^{tst} \}$. Using 2-fold cross-validation, we train and test the proposed G-GAN DBMP model along a linear SVM classifier to label the multiplex $\mathbb{\hat{M}}$ into male or female. The training of the SVM classifier is proceeded by a feature selection step using the training samples. We particularly use infinite feature selection (IFS) \cite{roffo2015} to select the most reproducible features distinguishing between male and female brain multiplexes. 

\textbf{Gender-related brain multiplex connectivity discovery.} Following feature selection, we identify the top $n_f$ gender-discriminative features. We design a feature scoring algorithm by quantifying feature reproducibility across validation folds depending on the occurrence of each selected feature and its weight given by the IFS algorithm. Indeed, the more frequently a feature appears in the top $n_f$ set, the higher its score is.

\section{Results and Discussion}

\begin{figure}[!htpb]
\centering
\includegraphics[width=12cm]{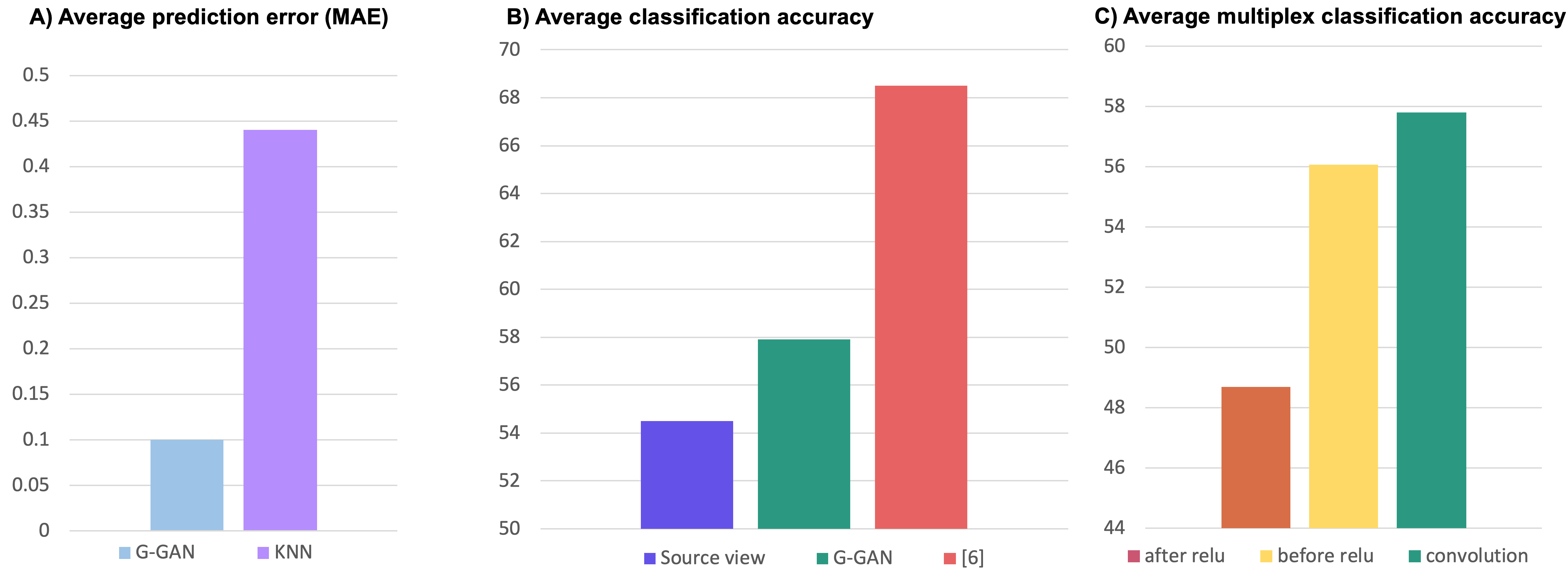}
\caption{\emph{Average classification accuracy using linear SVM classifier and multiplex prediction error.}  \textbf{(A)} We report the the mean absolute error (MAE) between the predicted and ground truth target intra-layers by (i) the proposed G-GAN method and (ii) KNN algorithm. \textbf{(B)} Average classification accuracy using: (i) a single source intra-layer network, (ii) the predicted multiplex, and  (iii) the ground-truth multiplex \cite{raeper2018a}. \textbf{(C)}  Predicted multiplex classification accuracy while defining the multiplex inter-layer as: (i) the learned G-GAN inter-layer preceding relu, (ii) the learned G-GAN inter-layer following relu, and (iii) the convolutional layer between both intra-layers as in \cite{raeper2018a}.}
\label{fig:2}
\end{figure}

\begin{figure}[!htpb]
\centering
\includegraphics[width=12cm]{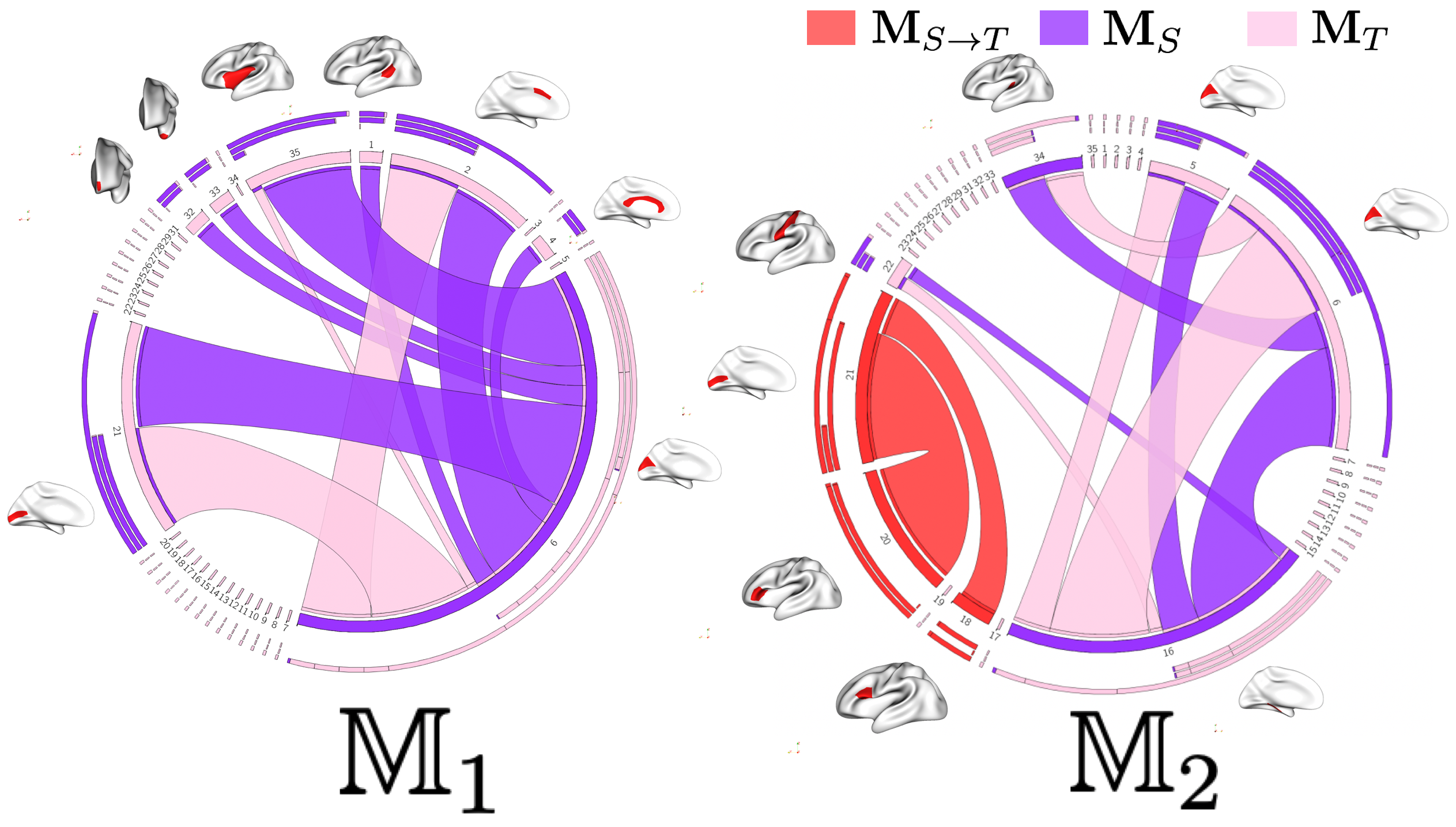}
\caption{\emph{Discovery of the top gender discriminative multiplex connectivities.}  We display the top 10 most discriminative multiplex connections for $\mathbb{M}_1$ ($S =$ maximum principal curvature MBN and $T=$  mean cortical thickness MBN) and $\mathbb{M}_2$ ($S =$ mean average curvature and $T=$  mean cortical thickness).}
\label{fig:2}
\end{figure}

\textbf{Evaluation dataset and method parameters.}  We evaluated our proposed framework on a dataset of 400 healthy subjects (226 females with mean age {$= (21.6 \pm 0.8)$} and 174 males {with mean age $= (21.6 \pm 0.9$}) from the Brain Genomics Superstruct Project \cite{Buckner:2012}.
T1-weighted images were acquired using a $1.2mm$ isotropic resolution. Test-retest reliability was established with a correlations range from $0.75$ for the estimated cortical thickness of the right medial prefrontal cortex to $0.99$ for the estimated intracranial volume. We used FreeSurfer processing pipeline to reconstruct the left and right cortical hemispheres. Then we parcellated each hemisphere into 35 regions using Desikan-Killiany atlas. For each subject, we created 4 morphological brain networks (MBN) of size {$35 \times 35$} as introduced in \cite{soussia2017high,mahjoub2018}, derived from the following cortical attributes respectively: cortical thickness, sulcal depth, mean average curvature, and maximum principal curvature.

\textbf{Method evaluation and comparison methods}. Due to the absence of state-of-the-art methods focusing on learning how to predict brain multiplexes using G-GAN as well as investigating how gender differences manifest in brain multiplexes, we resorted to comparing our framework against widely used methods such as KNN. This is a valid approach adopted by high-impact publication venues such as \cite{samusik2016}, where innovative proposed frameworks are benchmarked against the most commonly used techniques. KNN is a widely used algorithm due to its robustness to noisy data. Basically, for each testing subject with a source network, we first identify its top $K$ most similar training source brain networks. Next, we retrieve their corresponding target networks, which we average to predict the target testing intra-layer. To evaluate the potential of the predicted multiplex in boosting gender classification, we benchmarked our predicted multiplexes by G-GAN against classification using: (i) solely source intra-layer (i.e., from a single network), and (ii) ground truth shallow multiplexes  where multiplexes were generated using two MBNs and one convolutional layer between them as in \cite{raeper2018a,lisowska2017}.

\textbf{Fig}~\ref{fig:2}--A shows the mean absolute error (MAE) between the predicted target intra-layers and the ground truth ones. For KNN, we report the average performance when varying the number of top $K$ selected neigbnors from 2 to 10. Notably, G-GAN predicted multiplexes significantly outperforms the baseline KNN method. \textbf{Fig}~\ref{fig:2}--B displays classification results averaged across multiplexes and number of selected features $n_f \in \{ 310, \dots, 350 \}$ using three methods: source MBN, predicted multiplex by G-GAN, and ground-truth multiplexes. The predicted multiplexes significantly boosted the classification results in comparison to solely using the ground-truth source MBN. We also notice that it is quite close to the ground-truth multiplex. This classification experiments present an indirect method for evaluating the quality and reliability of the predicted multiplexes.

\textbf{Insights into learning the multiplex inter-layer.} Instead of computing the inter-layer by convolving both intra-layers, we extracted the \emph{learned} final inter-layer in the encoding part of the translator: before and after the rectified linear unit (ReLU) activation function.  \textbf{Fig}~\ref{fig:2}--C displays the average classification accuracy using the learned embedding of the source intra-layer as its gets translated into the target intra-layer (i.e., a hybrid inter-layer bridging the gap between the source and target intra-layers) and the conventional convolutional inter-layer \cite{raeper2018a}. Although, this experiment is considered as the first attempt to learn the multiplex inter-layer in the state-of-the-art, the convolutional predefined inter-layer still achieved the best classification result. This might be explained that the learned inter-layer is learned without any supervision by the sample class labels.

\textbf{Gender multiplex fingerprint discovery.} As a proof of concept, \textbf{Fig}~\ref{fig:3} displays results from two multiplexes $\mathbb{M}_1$ and   $\mathbb{M}_2$, where we report the top $10$ most discriminative multiplex connectional biomarkers for distinguishing between male and female morphological brain networks. These connectional fingerprints were captured on both low and high-order levels. For instance, in $\mathbb{M}_1$ ($S =$ maximum principal curvature MBN and $T=$ mean cortical thickness MBN), we found that the most gender-discriminative connectivity links (entorhinal cortex $\leftrightarrow$ pericalcarine cortex) in both intra-layers. In fact, the entorhinal cortex is known to be correlated with spatial memory processing, where \cite{cherney2008} showed that male participants outperform female participants. Besides, the pericalcarine cortex controlling visual processing and spatial navigation was identified as a gender discriminative region in \cite{ingalhalikar2014}. While in the multiplex $\mathbb{M}_2$ ($S =$ mean average curvature and $T=$ mean cortical thickness), we found that the most discriminative connectivity included the ( pars triangularis $\leftrightarrow$ pericalcarine cortex). For instance, the pars triangularis is known to be related to language and speech processing, which is in line with the observation that females are better at language processing than males as reported in \cite{bourne2005}. This hallmark connectivity was found in the convolutional inter-layer. We also notice other relevant brain multiplex connectivities distributed across intra-layers. In our future work, we will investigate the reproducibility of the discovered multiplex biomarkers fingerprinting gender using different perturbation strategies of the training set while tracking the reproducibility of the top ranked features \cite{Georges:2020}.

\section{Conclusion}

In this paper, we presented the first framework for brain multiplex synthesis from a source brain network with application to gender classification. Specifically, we rooted our method in geometric generative adversarial network (G-GAN) nicely preserving local and global brain connectional patterns, where a network translator and discriminator are adversarially trained. The predicted multiplexes significantly boosted the classification results compared to using single source networks. More importantly, we identified for the first time both low-order and high-order morphological brain connectivities encoding gender differences using large GSP dataset. There are many possible future directions yet to explore, such as learning \emph{discriminative} multiplex inter-layers by integrating a second conditional discriminator by class labels.

\section{Supplementary material}

We provide two supplementary items for reproducible and open science:

\begin{enumerate}
	\item A 6-mn YouTube video explaining how our prediction framework works on BASIRA YouTube channel at \url{https://youtu.be/iTjPtC4BULc}.
	\item An improved version of the adversarial brain multiplex generation code is available on GitHub at \url{https://github.com/basiralab/ABMT}. 
\end{enumerate}

\section{Acknowledgement}

This project has been funded by the 2232 International Fellowship for
Outstanding Researchers Program of TUBITAK (Project No:118C288, \url{http://basira-lab.com/reprime/}) supporting I. Rekik. However, all scientific contributions made in this project are owned and approved solely by the authors.

\bibliography{Biblio3}
\bibliographystyle{splncs}
\end{document}